\documentclass[aps,prl,preprint,draft,groupedaddress,amssymb,showpacs]{revtex4}
\usepackage{epsf}
\usepackage{dcolumn}
\begin{document}
\title{Evolution of the electronic structure with size in II-VI semiconductor
nanocrystals}
\author{Sameer Sapra}
\author{D. D. Sarma}
\altaffiliation[Also at ] {Jawaharlal Nehru Centre for Advanced
Scientific Research, Bangalore and Centre for Condensed Matter Theory, IISc.}
\email{sarma@sscu.iisc.ernet.in} \affiliation{Solid State and
Structural Chemistry Unit, Indian Institute of Science,
Bangalore-560012, India}
\begin{abstract}
In order to provide a quantitatively accurate description of the
band gap variation with sizes in various II-VI semiconductor
nanocrystals, we make use of the recently reported tight-binding
parametrization of the corresponding bulk systems. Using the same
tight-binding scheme and parameters, we calculate the electronic
structure of II-VI nanocrystals in real space with sizes ranging
between 5 and 80~{\AA} in diameter. A comparison with available
experimental results from the literature shows an excellent
agreement over the entire range of sizes.
\end{abstract}
 \maketitle

\section{Introduction}
It is now possible to grow a large variety of semiconductor
nanocrystals and also control their sizes to obtain mono-dispersed
particles.~\cite{murrayARMS00,alivisatosJPC96} A large number of
II-VI~\cite{murrayJACS93,vossmeyerJPC94,nandaCM00} and
III-V~\cite{micicJPC94,guzelianAPL96,guzelianJPC96} semiconductor
nanocrystals have been prepared over the past two decades. These
quantum dots are good candidates for electronic and optical
devices~\cite{gapoenkoBOOK98,woggonBOOK97,kastnerPT93,alivisatosSCI96,brusJPCS98}
due to their reduced dimensions, enabling one to reduce the size
of electronic circuitry. Also, due to the increased oscillator
strengths in these nanocrystals as a result of quantum
confinement,~\cite{brusJPC86} these are expected to have higher
quantum efficiencies in applications such as light emission. This
is a direct consequence of a greater overlap between the electron
and the hole wavefunctions upon size reduction. Moreover, one can
tune these properties to suit a specific application by merely
changing the size of the nanocrystals. For example, the band gap
of CdSe can be varied from 1.9~eV to 2.7~eV by changing the size
of the particle from 5.5~nm to 2.3~nm.~\cite{dabbousiJPC97}
Alongwith the band gap of the particle, the photoluminescence can
also be varied through the red to the blue region of the visible
spectrum.~\cite{dabbousiJPC97} This quantum size effect can be
explained qualitatively by considering a particle-in-a-box like
situation where the energy separation between the levels increases
as the dimensions of the box are reduced. Thus, one observes an
increase in the band gap of the semiconductor with a decrease in
the particle size

On a more quantitative footing, various different theoretical
approaches have been employed to account for the variation in the
electronic structure of nanocrystallites as a function of its
size. The first explanation for the size-dependence of electronic
properties in nanocrystals was given by Efros and
Efros.~\cite{efrosSPS82} It is based on the effective masses of
the electron ($m_e^*$) and the hole ($m_h^*$). Known as the
Effective Mass Approximation (EMA), it is solved by taking various
choices for the electron and hole wavefunctions and solving the
effective mass equation variationally. In most EMA calculations,
the confining potentials for the electron and the hole have been
assumed
infinite.~\cite{brusJPC86,efrosSPS82,brusJCP84,kayanumaSSC86,shmidtCPL86,nairPRB87}
Therefore, the electron and the hole wavefunctions vanish at and
beyond the surface of the nanocrystal, without the possibility of
any tunnelling. In the strong confinement regime, where $R$, the
nanocrystal radius, is much smaller than $a_B$, the Bohr exciton
radius, Brus proposed~\cite{brusJCP84} the following expression
for the bandgap of the finite sized system
\begin{widetext}
\begin{equation}
E(R)~=E_g~+~\frac{\hbar^2}{2}~(\frac{1}{m_e^*}~+~\frac{1}{m_h^*})~\frac{\pi^2}{R^2}
~-~1.786~\frac{e^2}{\epsilon R}~-~0.248~E^*_{Ry}
\end{equation}
\end{widetext}

\noindent where, $E_g$ is the bulk band gap. The second term is
the kinetic energy term containing the effective masses, $m_e^*$
and $m_h^*$, of the electron and the hole, respectively. The third
term arises due to the Coulomb attraction between the electron and
the hole, and the fourth term due to the spatial correlation
between the electron and the hole which is generally small
compared to the other two terms.

EMA calculations have also been reported where a finite confining
potential was used to account for the passivating agents that coat
the surface of the nanocrystals in order to arrest their growth.
Finite potential calculations are shown to improve the description
for CdS nanocrystals to a large extent.~\cite{kayanumaPRB90}
Another improvement to the single band EMA is the inclusion of
multiple bands for describing the hole effective mass. This is
prompted by the fact that the top of the valence band for II-VI
semiconductors is comprised of triply degenerate bands at the
$\Gamma$ point and thus is better defined using a multi-band
theory. To account for this degeneracy,
Einevoll~\cite{einevollPRB92} and Nair {\it et
al.}~\cite{nairPRB92} have used the effective bond-orbital model
for the hole wavefunction, while the electron is described by a
single-band EMA. Finite barrier heights and the electron-hole
Coloumb attraction are included in the calculation and exciton
energies are obtained variationally in an iterative Hartree
scheme. The multi-band and finite potential EMA methods explain
the experimental results reasonably well, but lack the predictive
capabilities desirable of a theoretical model, since the finite
potentials need to be adjusted to match the experimental results
in each specific case. Psuedopotential calculations have also been
carried out to study the variation of electronic structure with
the nanocrystal size.~\cite{ramaPRL91,wangPRB95,fuPRB97} Recently,
the semi-empirical pseudopotential method has been employed to
calculate the electronic structure of Si, CdSe~\cite{wangPRB95}
and InP~\cite{fuPRB97} nanocrystals. The atomic pseudopotentials
are extracted from first principles LDA calculations on bulk
solids. Thus, the wave functions are LDA-like while the band
structures, effective masses, and deformation potentials are made
to match experimental results. This method provides a reasonable
description of the electronic structure of the nanocrystals.
However, major computational efforts and difficulties do not allow
one to calculate the properties of large sized nanocrystals.

The tight-binding (TB) scheme has been employed by a number of
researchers over the past
decade.~\cite{LLPRB89,leungPRB98,condePRB01,sapraJPD03,hillJCP93,hillJCP94,sapraNL02,allanAPL00}
This method enjoys several advantages over the other methods
discussed above, explaining its popularity. Compared to EMA, both
pseudopotential method and the tight-binding approach provide a
substantial improvement in the accuracy of the results. The
tight-binding method has the further advantage of being
significantly less demanding in terms of computational efforts,
besides providing a simple physical picture in terms of the atomic
orbitals and hopping interactions defined over a predetermined
range. A detailed analysis of the first-principle electronic
structure calculations can lead to a judicious tight-binding
scheme that is minimal in terms of the dimension of the
Hamiltonian matrix and yet is highly accurate due to the use of a
physical and realistic basis.~\cite{sapraPRB02} The earliest such
TB parametrization was provided by Vogl {\it et
al.}~\cite{voglJPCS83} who used a TB model with the $sp^3s^*$
orbital basis in order to describe the electronic structure of
bulk semiconductors. The $s^*$ orbital was employed in an {\it ad
hoc} manner in addition to the $sp^3$ orbital basis in order to
improve the TB-fit to the $ab-initio$ band dispersions.
Subsequently, this TB model was used by Lippens and
Lannoo~\cite{LLPRB89} to calculate the variations in the band gap
for the corresponding semiconducting nanocrystals as a function of
the size. Though their results are in better agreement compared to
the infinite potential EMA, the $sp^3s^*$ TB model tends to
underestimate the band gap. The main problem with the $sp^3s^*$
model appears to be a failure to reproduce even the lowest lying
conduction band within that scheme.~\cite{voglJPCS83} Improvements
in the nearest neighbor $sp^3s^*$ model have been carried out by
including the spin-orbit coupling and the electron-hole
interaction.~\cite{leungPRB98,condePRB01} However, to account for
the conduction bands, the inclusion of $d$ orbitals becomes
necessary.~\cite{sapraJPD03,allanAPL00} This has been shown in the
case of InP~\cite{sapraJPD03} nanocrystals, a III-V semiconductor,
that TB model with the $sp^3d^5$ orbital basis for the anion and
the $sp^3$ basis for the cation with next nearest neighbor
interactions, for both the anion and the cation, gives excellent
agreement with the experimental data. In a recent work, we have
shown that the $sp^3d^5$ orbital basis for both the cation and the
anion and the inclusion of the next nearest neighbor interactions
for the anions provide a very good description of the electronic
structure of bulk II-VI semiconductors.~\cite{sapraPRB02} This
model is shown to describe accurately the band gap and the band
dispersions for both the valence and the conduction bands over the
energy range of interest. Therefore, this improved model and the
parametrization should provide a good starting point for
calculating the electronic properties of corresponding
nanocrystals, provided the model and parameters are transferable
from the bulk to the cluster limit. $Ab~initio$ calculations for a
CdS cluster of about 16 {\AA} diameter~\cite{note-transfer} as
well as results of Ref.~\cite{ramaPRL91} suggest that the present
scheme is of sufficient accuracy down to about 16 {\AA}, though
the applicability of this approach may be limited for still
smaller sized clusters. In order to explore the possibility of
utilizing it effectively, we have used this model for calculating
the band gap variation over a wide range of sizes for A$^{\rm
II}$B$^{\rm VI}$ semiconductor nanocrystals, with A~=~Cd or Zn and
B~=~S, Se or Te, comparing the calculated results with the
experimental data from the literature. The present results show a
good agreement with experimental results, where ever available.

\section{Theoretical Procedure}
The appropriate minimal TB model for the bulk electronic structure
of group II-VI semiconductors was developed in
Ref.~\cite{sapraPRB02} by analyzing the atomic wave-function
characters of the various bands. This established $sp^3d^5$ basis
with the cation-anion and anion-anion interactions as the suitable
model. The tight-binding electronic parameters, namely the orbital
energies and the hopping strengths, were determined by fitting the
$ab-initio$ band dispersions to the band dispersions obtained from
the tight-binding Hamiltonian, given by
\begin{widetext}
\begin{equation}
H~=~\sum_{il_1\sigma}
\epsilon_{l_1}~a^{\dagger}_{il_1\sigma}~a_{il_1\sigma}
+~\sum_{ij}\sum_{l_1l_2\sigma}~(t_{ij}^{l_1l_2}~a^{\dagger}_{il_1\sigma}~a_{jl_2\sigma}~+~h.c.)
\end{equation}
\end{widetext}
\noindent where, the electron with spin ${\sigma}$ is able to hop
from the orbitals labelled $l_1$ with onsite energies equal to
$\epsilon_{l_1}$ in the $i^{th}$ unit cell to those labelled $l_2$
in the $j^{th}$ unit cell, with a hopping strength
$t_{ij}^{l_1l_2}$; the summations $l_1$ and $l_2$ running over all
the orbitals considered on the atoms in a unit cell, and $i$ and
$j$ over all the unit cells in the solid. We use exactly the same
model with the parameter strengths given in Ref.~\cite{sapraPRB02}
to calculate the electronic structure of corresponding
nanocrystals as a function of the size.

We build the cluster shell by shell, starting from a central atom.
For the tetrahedrally coordinated compounds in the zinc-blende
structure, the central atom, say the cation, is surrounded by a
shell of four anions. In turn each of these anions is coordinated
by four cations, one of them being the central cation. The other
three cations form a part of the next shell. The clusters are
generated in this manner by successive addition of shells.
Assuming a spherical shape of the cluster, the diameter, $d$, is
given by
\begin{equation}
d~=~a~[\frac{3N}{4\pi}]^{\frac{1}{3}}
\end{equation}

\noindent where, $a$ is the lattice constant and $N$ the number of
atoms present in the nanocrystal. Table~I lists the number of
atoms present upto a given shell and the diameter of the
nanocrystal for various A$^{\rm II}$B$^{\rm VI}$ compounds. The
Hamiltonian matrix for any given sized cluster is obtained from
Eq.~2 with the same atomic orbital basis and electronic parameter
strengths as given in Ref.~\cite{sapraPRB02} and is diagonalized
to obtain the eigenvalue spectra for the nanocrystal. Direct
diagonalization methods are practical only for cluster sizes
containing less than $\sim$1500 atoms. For larger clusters, we use
the Lanczos iterative method.~\cite{dagottoRMP94}

The Lanczos algorithm uses a starting basis function $\mid\phi_0>$
which can be a linear combination of the atomic orbitals,
$\varphi_i$'s, {\it i.e.}
$$ \mid\phi_0>~=~\sum_i~c_i\mid\varphi_i> $$

\noindent Once the starting basis function has been generated, a new basis function
$\mid\phi_1>$ is generated by applying the Hamiltonian and then making the resulting
function orthogonal to $\mid\phi_0>$.
$$ \mid\phi_1>~=~H\mid\phi_0>~-~\frac{<\phi_0{\mid}H\mid\phi_0>}{<\phi_0\mid\phi_0>}
\mid\phi_0>$$

\noindent Then onwards the subsequent basis functions can be
generated by using the recursion formula
\begin{widetext}
$$ \mid\phi_{n+1}>~=~H\mid\phi_n>~-~a_n\mid\phi_n>~-~b_n^2\mid\phi_{n-1}>~~~~~~n~=~0,1,2,...$$
\end{widetext}

\noindent where,~~~~$a_n$~=~$\frac{<\phi_n{\mid}H\mid\phi_n>}{<\phi_n\mid\phi_n>}$~,~
$b_n^2$~=~$\frac{<\phi_n\mid\phi_{n}>}{<\phi_{n-1}\mid\phi_{n-1}>}$ \\

\noindent with $b_0$~=~0 and $\mid\phi_{-1}>$~=~0. By
construction, each basis function is orthogonal to the previously
generated basis functions. Here the $a_n$'s are the diagonal
elements, while $b_n$'s are the off-diagonal terms of the
Hamiltonian matrix. Diagonalization of this tridiagonal matrix is
less time consuming and gives the eigenvalue spectrum for the
clusters. We choose the starting seed vector to be a particular
orbital of an atom. The eigen spectrum thus obtained is composed
of only those orbitals that couple with the seed vector. Thus,
taking each orbital of every atom in the cluster we obtain the
entire density of states. Due to the underlying symmetry in the
nanocrystal we need not perform calculations for all atomic
orbitals as seed vectors, but only those with distinct symmetries.

The band gap for a particular sized nanocrystal is then calculated
by subtracting the energy of the top of the valence band (TVB)
from that of the bottom of the conduction band (BCB). However, the
determinations of the TVB and the BCB become ambiguous due to the
presence of dangling bonds at the surface of the nanocrystals.
These non-bonded states lie in the band gap region of the
nanocrystals. These surface states need to be either selectively
disposed off~\cite{LLPRB89} or
passivated~\cite{hillJCP93,hillJCP94} in order to remove the
midgap states. Once the surface states are passivated, the band
gap can be easily determined. In the present work, we have
passivated the surfaces of the nanocrystals in order to remove the
mid-gap states from the calculations.

\section{Results and Discussion}
The various steps involved in the calculations for the variation
of the band gap with size are quite similar for the different
A$^{\rm II}$B$^{\rm VI}$ compounds studied here. We therefore use
the case of CdS as an example to illustrate all the steps and
various considerations, prior to presenting comprehensive results
for all the systems together at the end.

Fig.~\ref{doscomparison}a shows the first-principle results for
the density of states (DOS) for CdS bulk obtained from the
linearized muffin-tin orbital (LMTO) method with the atomic sphere
approximation (ASA). It should be noted that the parameters
appearing in the TB Hamiltonian (Eq.~1) were
determined~\cite{sapraPRB02} by a least-squared-error approach in
order to obtain dispersions at high symmetry points and a few
other $k$-points along the symmetry directions in the Brillouin
zone. Since we are eventually interested in the density of states
which involves an integration over the entire momentum space, we
have explicitly verified in each case that the DOS calculated
within the TB approach is very similar to the one obtained from
the LMTO-ASA method. We illustrate this point with the help of DOS
calculated within the TB model for CdS with parameter strengths
from Ref.~\cite{sapraPRB02}; this TB DOS is shown in
Fig.~\ref{doscomparison}b with the same energy scale as in
Fig.~\ref{doscomparison}a. We notice an excellent agreement of the
TB DOS with the LMTO DOS over the entire range of the energy
considered.

While the electronic structure of small sized nanocrystals is
known to be pronouncedly dependent on the size, larger sized
nanocrystals are expected to resemble the bulk in terms of their
electronic structures; evidently in the limit of the large size,
the electronic structure of the nanocrystal must smoothly evolve
into that of the bulk. It is known that the quantum confinement
effect is generally small for a nanocrystal with typical size
larger than the excitonic radius. The excitonic diameter of CdS is
about 58~{\AA}.~\cite{yoffeAP93} We consider a CdS cluster of
76~{\AA} containing 9527 atoms that is considerably larger than
the excitonic diameter. In Fig.~\ref{doscomparison}c we show the
DOS for this large CdS cluster. The DOS of the nanocrystal indeed
resembles the bulk DOS closely, as is evident in
Fig.~\ref{doscomparison}, apart from the discrete nature of the
DOS arising from the finite size of the nanocrystallite system.

As discussed in the previous section, the dangling orbitals on the
surface atoms appear within the band gap region, complicating the
identification of the band gap.  Fig.~\ref{znsovercds}a shows the
normally obtained DOS for a 46~{\AA} CdS nanocrystal; the
corresponding inset shows an expanded view of the band gap region.
As one can clearly see in the expanded view, there are many states
spread out over an energy range appearing between the valence band
and the conduction band due to the aforementioned dangling bonds
within the band gap region. As already discussed, different
authors approached the problem of dangling bonds or its removal
from the DOS in different ways. For example, Lippens and
Lannoo~\cite{LLPRB89} got rid of the dangling bonds by removing
the unconnected orbitals on the surface atoms in order to obtain
the band gap free of the mid-gap states. In spirit, this approach
is similar to the infinite potential barrier on the surface of the
nanocrystal assumed in the infinite potential EMA. Akin to the
finite potential EMA, we choose to passivate the surface with a
layer of atoms, whose electronic parameters are so chosen that the
hopping interactions between the surface atoms and the passivating
atoms are stronger compared to those in the bulk of the
nanocrystal. Specifically, we choose only the $s$ orbital basis on
the passivating atoms with the tight-binding hopping parameters
about 2-3 times larger than that of A-B interactions. The
corresponding DOS of the passivated nanocrystals of CdS is shown
in Fig.~\ref{znsovercds}b. In the main frame of the figures, the
unpassivated case in the upper panel and the passivated case in
the lower panel appear almost identical, suggesting that the
intrinsic electronic structure of the nanocrystals remain largely
unaffected by the passivation. In order to illustrate the effect
of passivation on the midgap states, we show an expanded view of
the band gap region between the TVB and the BCB in the inset to
Fig.~\ref{znsovercds}b. This inset shows that the surface
passivation is indeed effective in removing the midgap states,
present in the inset to Fig.~\ref{znsovercds}a, illustrating the
unpassivated case.

Most often, the total band gap variation as a function of the size
of the nanocrystal is reported in the literature.~\cite{notetoteg}
This is primarily motivated by the fact that this quantity,
$\Delta E_g$, is easily determined by experimental UV-visible
absorption spectroscopy, that is a routine characterization tool.
However, it is to be noted that the total change in the band gap
of any material is simultaneously contributed by shifts of the
valence and the conduction band edges away from each other. In
general, the shift of the top of the valence band is not the same
as that of the bottom of the conduction band. Moreover, there are
recent studies, though few in
number~\cite{colvin91,luning99,vanbuuren98,nandaTHESIS} that
report the individual shifts in TVB and BCB as a function of the
size employing various forms of high energy spectroscopies, such
as the photoemission and the x-ray absorption spectroscopies.
Thus, it is desirable to compute these shifts of the individual
band edges with the size of the nanocrystallite. The variation of
TVB (circles) and the BCB (squares) with respect to the bulk
values are calculated for different sized passivated nanocrystals
and shown in Fig.~\ref{cdsbandedge}. As expected, the shifts of
the band edges decrease smoothly to zero for large sized
nanocrystals in every case. We find that the shift in the BCB is
in general much larger compared to the shift in the TVB for any
given size of the nanocrystal; this indicates that the shifts in
the total band gap as a function of the nanocrystal size are
always dominated by the shifts of the conduction band edge in
these systems. A larger shift for the BCB is indeed expected in
view of the fact that the band edge shifts are related inversely
to the corresponding effective masses (see Eq.~1) and the
effective mass of the electron is always much smaller than that of
the hole in these II-VI semiconductors. For example, $m_e^*$ and
$m_h^*$ in CdS are 0.18 and 0.53, respectively.

In the spirit of EMA, one can attempt to describe the shifts in
the conduction and valence band edges, as
\begin{equation}
\Delta E^{edge}_i~=~\frac{a_i}{d^{b_i}}
\end{equation}
where  $\Delta E^{edge}_i$ is the variation in the band edge with
diameter $d$; $i~=~h$ for TVB and $i~=~e$ for BCB. Comparing with
the EMA (Eq.~1), one expects the fitting parameter $a_i$ to be
inversely proportional to the electron (for BCB, $i~=~e$) or hole
(for TVB, $i~=~h$) effective mass and $b_i$ to equal 2. We have
fitted the shifts in BCB and TVB as a function of $d$ with Eq.~4
by varying the parameters $a_i$ and $b_i$ within a
least-squared-error approach; the resulting best fits are shown in
Fig.~\ref{cdsbandedge} by the solid lines overlapping the
calculated data points. We find that the fits are reasonable,
though not very good, in most cases. More importantly, these fits
suggest a gross deviation from the EMA predictions; for example,
the variations in TVB and BCB shown in Fig.~\ref{cdsbandedge} are
far from the EMA-like $d^{-2}$ dependence and instead the best
exponent for $d$ is  in the range of 1.13-1.27 for TVB and
0.65-1.05 for BCB, as shown in Table~II.

Fig.~\ref{all} shows the variation of the shift in the band gap
($\Delta E_g$) for the A$^{\rm II}$B$^{\rm VI}$ semiconductor
nanocrystals with A=Zn, Cd and B=S, Se and Te as a function of the
nanocrystal size. $\Delta E_g$ is calculated in the present model
after subtracting the Coulomb term (third term of Eq.~1) from the
calculated difference between the TVB and the BCB to account for
the excitonic binding energy, since the experimental data obtained
from the UV-absorption include the contribution from the excitonic
binding energy. The solid line passing through the calculated data
points (small solid circles) is the best fit to the calculations.
The best fit is obtained by using simple exponential functions
relating $\Delta E_g$ to the diameter of the nanocrystallites as
\begin{equation}
\Delta~E_g~=~a_1~e^{-d/b_1}~+~a_2~e^{-d/b_2}
\end{equation}
While this expression is entirely phenomenological, it has the
correct limiting behavior at large $d$. The advantage of such a
best fit is that the $\Delta E_g$ for any given system can be
readily calculated for any size of the nanocrystallites with the
knowledge of the parameter values $a_1$, $b_1$, $a_2$ and $b_2$,
which are tabulated in Table~III for all the systems investigated
here. For comparison, we also show in the same panels the results
obtained from the $sp^3s^*$ nearest neighbor TB model (dashed
line)~\cite{LLMSE91} and the results from the EMA equation (dotted
line)~\cite{madelungBOOK92}. Experimental results available in the
literature are also plotted as scattered points with different
symbols for comparison with the calculated
results.~\cite{vossmeyerJPC94,nandaCM00,nakaokaLANG97,rossettiJCP85,inoueLANG94,
yanagidaJPC90,quinlanLANG00,hinesJPC98,junCC01,wangPRB90,
torimotoJPC01,nandaPRB99,dabbousiCM94,rogachJPC99,gorerJPC94,mastaiJPC97,
masumotoPRB97,arizpeJPCS00} There is a plethora of experimental
data for ZnS, CdS, CdSe and CdTe and we see that the present
approach provides a better description of the experimental data in
all these cases. The case of ZnSe, where the experimental results
are limited, also exhibits good agreement between the experiment
and the theory. In the case of ZnTe, the present as well as the
earlier calculations give almost similar descriptions;
unfortunately, the experimental data are limited and there are
large uncertainties in the data, so it is difficult to compare the
experimental results with our calculations. For most of the cases,
the $sp^3d^5$ model with the next nearest neighbor interactions is
in better agreement with the experiments compared to the
nearest-neighbor-only $sp^3s^*$ model. This is due to the fact
that the sole $s^*$ orbital does not account well for the
unoccupied states. These can only be described by the inclusion of
the empty anionic $d$ orbitals and the anion-anion interactions
which are of significance in the description of the bulk
electronic structure.~\cite{sapraPRB02}

\section{Conclusions}
We have calculated the electronic structure as a function of the
nanocrystallite size for A$^{\rm II}$B$^{\rm VI}$ semiconductors
with A=Zn and Cd and B=S, Se and Te, using the tight binding
method with the $sp^3d^5$ orbital basis set including the A-B and
B-B interactions. It is shown that the shift in the top of the
valence band as well as that in the bottom of the conduction band
are different from the predictions based on the effective mass
approximation, not only in quantitative terms, but also
qualitatively. The calculated variations in the band gaps over a
wide range of sizes are compared with all experimental data
published so far in the literature. This comparison shows a very
good agreement in every case, suggesting the reliability and the
predictive ability of the present approach.

\section{Acknowledgements}
The authors thank O. K. Andersen and O. Jepsen for the LMTO codes
and P. Mahadevan for providing unpublished $ab~initio$ results for
$\sim$~16~{\AA} CdS nanocrystals. We acknowledge financial support
from the Department of Science and Technology, Government of
India.

\begin{table*}
\caption{The unit cell edge length for zinc blende phase ($a$),
number of shells ($n_s$), number of atoms ($N$) in $n_s$, and the
average diameter, $d$, for various A$^{\rm II}$B$^{\rm VI}$ semiconductors studied. }
\label{tablesize}
\begin{center}
\begin{tabular}{rrrrrrrr}
\hline
&&&&&&&\\
&&~~~~~~ZnS&~~~~~~ZnSe&~~~~~~ZnTe&~~~~~~CdS&~~~~~~CdSe&~~~~~~CdTe\\[0.5ex]
\multicolumn{2}{c}{$a$~({\AA})}&5.41&5.67&6.10&5.82&6.05&6.48\\[0.5ex]
\hline
&&&&&&&\\
$~~~n_s$ &~~~~~~$N$& \multicolumn{6}{c}{$d$({\AA})}\\[0.5ex]
\hline
3&17&8.63&9.04&9.73&9.28&9.65&10.34\\
4&41&11.57&12.13&13.05&12.45&12.94&13.86\\
5&83&14.64&15.34&16.51&15.75&16.37&17.53\\
6&147&17.71&18.56&19.97&19.05&19.81&21.22\\
7&239&20.83&21.83&23.48&22.41&23.29&24.95\\
8&363&23.94&25.09&26.99&25.76&26.77&28.68\\
9&525&27.07&28.38&30.53&29.13&30.28&32.43\\
10&729&30.20&31.66&34.06&32.49&33.78&36.18\\
11&981&33.35&34.95&37.60&35.87&37.29&39.94\\
12&1285&36.49&38.24&41.14&39.25&40.80&43.70\\
13&1647&39.63&41.54&44.69&42.64&44.32&47.47\\
14&2071&42.78&44.83&48.23&46.02&47.84&51.24\\
15&2563&45.93&48.14&51.79&49.41&51.36&55.01\\
16&3127&49.08&51.44&55.34&52.80&54.88&58.78\\
17&3769&52.23&54.74&58.89&56.19&58.41&62.56\\
18&4493&55.38&58.04&62.44&59.58&61.93&66.33\\
19&5305&58.53&61.35&66.00&62.97&65.46&70.11\\
20&6209&61.68&64.65&69.55&66.36&68.98&73.88\\
21&7211&64.84&67.95&73.11&69.75&72.51&77.66\\
22&8315&67.99&71.26&76.66&73.14&76.03&81.44\\
23&9527&71.15&74.57&80.22&76.54&79.56&85.22\\
\hline
\end{tabular}
\end{center}
\end{table*}

\begin{table*}
\caption{The values of the parameters $a$ and $b$ and used in
Eq.~4 for all the A$^{\rm II}$B$^{\rm VI}$ semiconductors
studied.} \label{tableuniv}
\begin{center}
\begin{tabular}{rrrrrrr}
\hline
&&&&&&\\
&~~~~~~ZnS&~~~~~~ZnSe&~~~~~~ZnTe&~~~~~~CdS&~~~~~~CdSe&~~~~~~CdTe\\[0.5ex]
\hline
$a_e$&15.72&13.71&8.23&24.47&24.43&16.38\\
$b_e$&1.01&0.91&0.65&1.05&1.05&0.92\\
$a_h$&-14.93&-13.31&-20.47&-7.76&-19.49&-19.03\\
$b_h$&1.18&1.15&1.13&1.27&1.19&1.13\\
\hline
\end{tabular}
\end{center}
\end{table*}

\begin{table*}
\caption{The values of the parameters $a$ and $b$ and used in
Eq.~5 for all the A$^{\rm II}$B$^{\rm VI}$ semiconductors
studied.} \label{tableall}
\begin{center}
\begin{tabular}{rrrrrrr}
\hline
&&&&&&\\
&~~~~~~ZnS&~~~~~~ZnSe&~~~~~~ZnTe&~~~~~~CdS&~~~~~~CdSe&~~~~~~CdTe\\[0.5ex]
\hline
$a_1$&7.44&2.65&5.10&2.83&7.62&5.77\\
$b_1$&2.35&7.61&10.35&8.22&6.63&8.45\\
$a_2$&3.04&1.90&1.05&1.96&2.07&1.33\\
$b_2$&15.30&23.50&97.93&18.07&28.88&43.73\\
\hline
\end{tabular}
\end{center}
\end{table*}

\clearpage
\begin{figure}
\caption{\label{doscomparison} Comparison of (a) LMTO DOS, (b) TB
DOS, and (c) DOS of a 76.5~{\AA} CdS nanocrystals.}
\end{figure}
\begin{figure}
\caption{\label{znsovercds} The DOS for 46~{\AA} (a) unpassivated
and (b) passivated CdS nanocrystals. The inset shows the expanded
region encompassing the top of the valence band and the bottom of
the conduction band, showing the removal of the midgap states when
the nanocrystal is passivated.}
\end{figure}
\begin{figure}
\caption{\label{cdsbandedge} The variation of the TVB and the BCB
with size for II-VI nanocrystals.}
\end{figure}
\begin{figure}
\caption{\label{all} The $sp^3d^5$ TB model  with the cation-anion
and anion-anion interactions (Ref.~\cite{sapraPRB02}, filled
circles) compared with the $sp^3s^*$ TB nearest neighbor model
(Ref.~\cite{LLMSE91}, dashed line) and the experimental data
points: (a) ZnS:~$\square$ Ref.~\cite{nakaokaLANG97}, $\triangle$
Ref.~\cite{rossettiJCP85}, $\nabla$ Ref.~\cite{inoueLANG94},
$\diamond$ Ref.~\cite{yanagidaJPC90}, $*$ Ref.~\cite{nandaCM00};
(b) ZnSe:~$\square$ Ref.~\cite{quinlanLANG00}, $\triangle$
Ref.~\cite{hinesJPC98}; (c) ZnTe:~$\square$ Ref.~\cite{junCC01}.
(d) CdS:~$\circ$ Ref.~\cite{vossmeyerJPC94}, $\triangle$
Ref.~\cite{wangPRB90}, $*$ Ref.~\cite{torimotoJPC01}, $\square$
Ref.~\cite{nandaPRB99}; (e) CdSe:~$\square$
Ref.~\cite{dabbousiCM94}, $\triangle$ Ref.~\cite{rogachJPC99}, $*$
Ref.~\cite{gorerJPC94}; (f) CdTe:~$\square$
Ref.~\cite{mastaiJPC97}, $\triangle$ Ref.~\cite{masumotoPRB97},
$*$ Ref.~\cite{arizpeJPCS00}. The solid line passing through the
calculated filled circles is the best fit to the calculated points
obtained using Eq.~5.}
\end{figure}
\end{document}